%% file: main.tex
\pdfoutput=1
\documentclass[final,5p,times,twocolumn,authoryear]{elsarticle}

\usepackage{amssymb}
\usepackage{amsmath}
\usepackage{booktabs}
\usepackage{graphicx}
\usepackage{microtype}
\input{numbers.tex}

\newcommand{\qlike}{QLIKE}
\newcommand{\RC}{\mathrm{RC}}
\newcommand{\tr}{\operatorname{tr}}

\journal{Results in Physics}

\begin{document}

\begin{frontmatter}

\title{Do econophysics-inspired inputs improve neural covariance forecasting? A leakage-free walk-forward study of intraday DJIA data}

\author[first]{Gabriele Ranco\corref{cor1}}
\ead{gabriele.ranco@alumni.imtlucca.it}
\cortext[cor1]{Corresponding author.}
\affiliation[first]{organization={IMT Institute for Advanced Studies},
            addressline={Piazza San Francesco 19},
            city={Lucca},
            postcode={55100},
            country={Italy}}

\begin{abstract}
Tools from econophysics---random matrix theory (RMT), Tsallis $q$-Gaussian tails, Hawkes self-excitation and diffusion-based regularisation---are often proposed as ingredients for financial risk models, but their incremental value over strong, simple benchmarks is rarely measured under strict temporal controls. We build a Physics-Inspired Neural Covariance model (PINC) that forecasts the next-day realised covariance matrix of \NStocks\ stocks of the New York Stock Exchange  from 1-min data (\NDays\ trading days, 2012--2015). The network uses RMT-cleaned correlations, a Tsallis tail index and Hawkes intensities as inputs, and may add a Fokker--Planck-inspired regulariser whose weight is chosen on validation data only. It is compared with sample, Ledoit--Wolf, RMT and extreme-value covariances, a HAR-RV model, an LSTM and a plain network of equal capacity, in a nested walk-forward design with \NFolds\ folds, \NTestDays\ out-of-sample days and \NSeeds\ seeds, verified by look-ahead and bit-identical leakage tests. The regulariser was rejected on validation in every fold, so PINC and the plain network coincide. The neural forecasts beat static estimators (Holm-adjusted $p<0.001$) but not HAR-RV ($p=\pHAR$) or the LSTM ($p=\pLSTM$); removing the RMT, Tsallis or Hawkes inputs has no detectable effect; and minimum-variance portfolios show no significant Sharpe-ratio gain after costs. In this setting the econophysics ingredients add no measurable value beyond a HAR-type benchmark.
\end{abstract}


\end{frontmatter}

\section{Introduction}
\label{sec:intro}

Estimating the covariance matrix of many assets from a limited sample is a central difficulty in portfolio construction. Minimum-variance allocation \citep{markowitz1952} amplifies the estimation error of the sample covariance, which has been addressed by shrinkage \citep{ledoitwolf2004a,ledoitwolf2004b} and by the spectral cleaning suggested by random matrix theory (RMT) \citep{laloux1999,plerou1999,plerou2002}. The physics literature offers further tools for financial time series: non-extensive $q$-Gaussian statistics for heavy tails \citep{tsallis1988}, self-exciting point processes for the clustering of extreme events \citep{hawkes1971,bacry2015}, and diffusion equations whose stationary solutions describe the distribution of volatility. A natural idea is to place such tools inside a flexible forecasting model and let the data decide how much they help.

Whether they do is an empirical question that requires two controls that are often missing. First, a benchmark of equal strength: the heterogeneous autoregression of realised volatility (HAR-RV) \citep{corsi2009} and recurrent networks \citep{hochreiter1997} are strong, inexpensive competitors, and a network trained with the physics-inspired term must be compared with the \emph{same} network trained without it. Second, strict temporal discipline: every estimator, scaler, hyper-parameter and early-stopping decision must be made without information that was not available at the decision date, and the claim must hold on data that were not used to choose anything. The recent enthusiasm for physics-informed neural networks \citep{raissi2019} concerns models whose loss contains the residual of a known partial differential equation; the regulariser used here is much weaker and does not enforce such a residual (Section~\ref{sec:pinc}).

We therefore ask six questions on the \NStocks\ stocksobserved every minute:
\begin{enumerate}
\item[Q1] Do RMT cleaning, the Tsallis tail index and Hawkes intensity improve the covariance forecast?
\item[Q2] Does the Fokker--Planck-inspired regulariser improve a plain network of equal capacity?
\item[Q3] Is the forecast better than static, HAR-type and recurrent benchmarks?
\item[Q4] Do minimum-variance portfolios built on it beat the benchmarks after transaction costs?
\item[Q5] Are value-at-risk (VaR) and expected-shortfall forecasts calibrated, and does the entropic VaR (EVaR) give different weights from minimum variance?
\item[Q6] Are the findings robust?
\end{enumerate}
The answer is largely negative, and this is our main result: the neural architecture improves on static estimators, but no incremental contribution of the econophysics ingredients can be detected against HAR-RV and the LSTM, and the physics-inspired regulariser is rejected by validation in every fold. The contribution of the paper is the controlled, reproducible protocol and this null finding, not a new winning model. Because the sample covers one market, one universe and \NTestDays\ out-of-sample days, the conclusions are restricted to this specification (Section~\ref{sec:discussion}).

\section{Data}
\label{sec:data}
\begin{figure*}[t]
\centering
\includegraphics[width=0.78\textwidth]{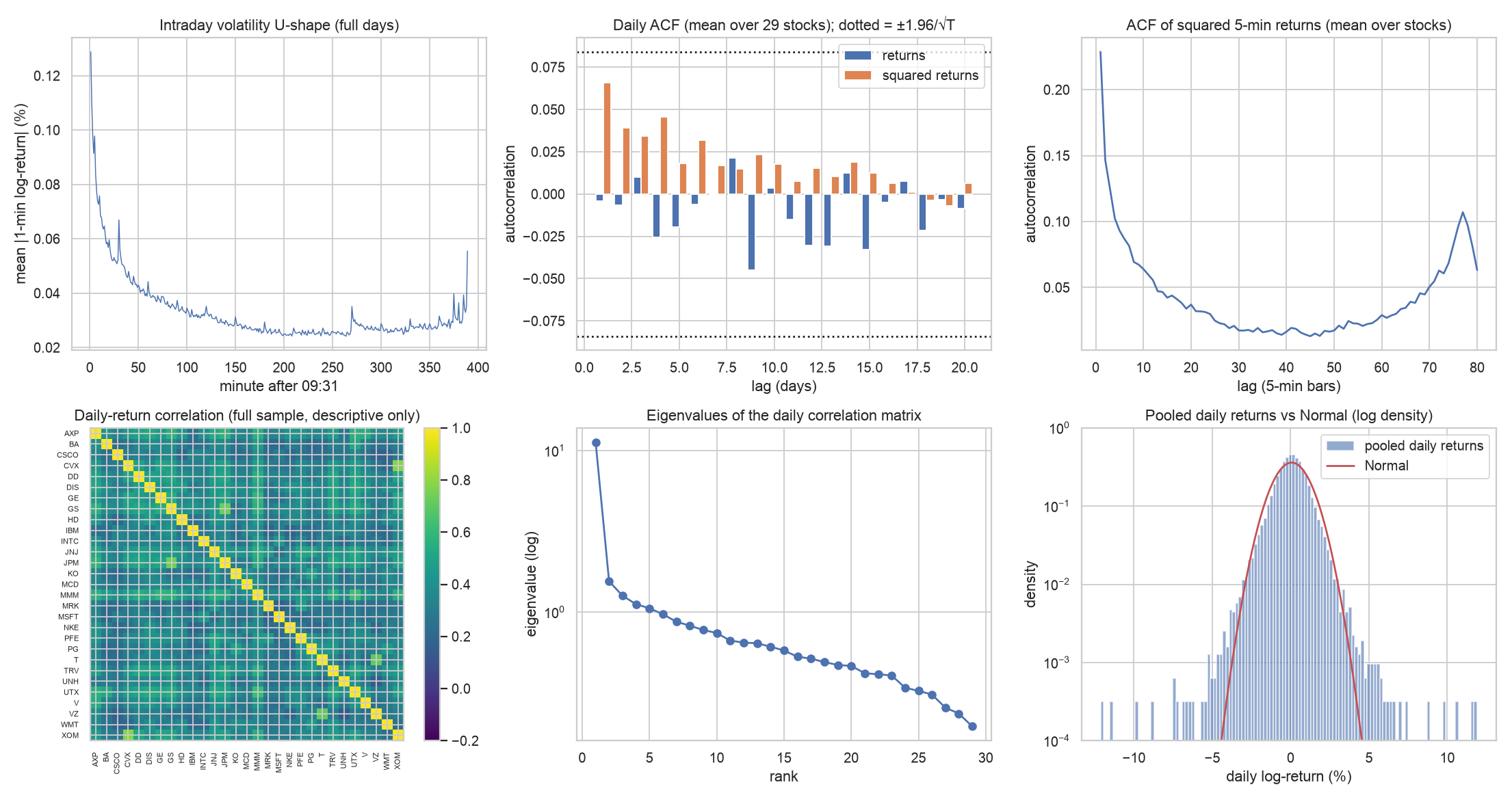}
\caption{Exploratory analysis of the data. Top: intraday volatility profile, autocorrelation of daily returns and squared returns, autocorrelation of squared 5-min returns. Bottom: full-sample daily correlation matrix (descriptive only; nothing is estimated on the full sample), its eigenvalues, and pooled daily returns against a Gaussian density.}
\label{fig:eda}
\end{figure*}

We consider a fixed universe of 29 individual stocks that were constituents of the Dow Jones Industrial Average (DJIA) at some point during the observation period. Their individual intraday price series were tracked over the study period from December 5, 2012, to February 4, 2015, using one-minute observations. The resulting dataset comprises 545 trading days. The selected universe is kept fixed throughout the empirical analysis to ensure consistency in the dimensionality of the covariance matrices.
We use 1-min bars of the \NStocks\ stocks from 2012-12-05 to 2015-02-04 (\NDays\ trading days, \NHalfDays\ of them half days closing at 13:00).. Official sessions are enforced (09:31--16:00, or 09:31--13:00 on half days) and late prints on half days are dropped. Missing bars are filled forward by at most two minutes and only within the same day, never across the overnight gap and never backward; this affects \PctImputed\,\% of the ticker-minutes, and the strict intersection of all stocks would otherwise discard \PctLost\,\% of the minutes. The daily returns computed with and without imputation are identical. Prices are sampled with the previous-tick rule and the realised covariance (RC) is the average over five 5-min grids shifted by one minute, a compromise that reduces microstructure noise without discarding data. Table~\ref{tab:data} summarises the sample and Fig.~\ref{fig:eda} shows the stylised facts: the intraday volatility U-shape, weak serial correlation of returns against persistent correlation of squared returns, a dominant market eigenvalue, and heavy daily tails.

\begin{table}[t]
\centering
\small
\input{tab_data.tex}
\caption{Data and experimental design. Strict intersection: all stocks have a bar in the minute.}
\label{tab:data}
\end{table}

\section{Methods}
\label{sec:methods}

\subsection{Forecasting target and timing}
\label{sec:timing}
Decisions are taken at the close of day $t$ using information up to and including that close. The target is the realised covariance matrix $\RC_{t+1}\in\mathbb{R}^{N\times N}$ of the next trading day, $N=\NStocks$, and a portfolio chosen at $t$ earns the close-to-close return of day $t+1$. The information, decision and realisation dates are stored separately and the notebook asserts that every information date precedes its realisation date. A negative control confirms that a deliberately mis-aligned target trips the assertion.

\subsection{Econophysics inputs}
\label{sec:inputs}

\emph{Random matrix theory.} For a window of $T=\corrWin$ daily returns the eigenvalues of a pure-noise correlation matrix lie below the Marchenko--Pastur edge $\lambda_+=(1+\sqrt{N/T})^2\approx\mpEdge$ \citep{marcenkopastur1967}. Eigenvalues not above $\lambda_+$ are replaced by their mean, which preserves the trace, and the matrix is renormalised to unit diagonal, giving the cleaned correlation $\Phi^{\rm RMT}_t$. On the data, the number of eigenvalues above $\lambda_+$ is on average \rmtSigMean\ (at most \rmtSigMax) against \rmtPerm\ when each column is shuffled independently, and the first eigenvalue carries on average \rmtShare\ of the variance.

\emph{Tsallis $q$-Gaussian.} For $1<q<3$ the $q$-Gaussian density is $f(x)\propto[1+(q-1)\beta(x-\mu)^2]^{-1/(q-1)}$. It is fitted by numerical maximum likelihood, not from the kurtosis, on daily returns. The median index is $q=\qMed$ (range \qMin--\qMax, no fit at a bound) and the $q$-Gaussian has a lower AIC than the Normal for \aicQNormal\ of \NStocks\ stocks. For $q>1$ the $q$-Gaussian is a re-parametrisation of the Student-$t$ law with $\nu=(3-q)/(q-1)$ degrees of freedom, so the likelihoods coincide and the comparison with the Student-$t$ is only a numerical check. The rolling index, estimated on past data only, is a network input.

\emph{Hawkes process.} Intraday jumps are identified with the Lee--Mykland test on seasonally standardised 5-min returns \citep{leemykland2008} (\hkNEvents\ events; daily jumps are also tested with the ratio statistic of \citealp{bns2006,huangtauchen2005}, \bnsJumpDays\ jump-days among \bnsTickerDays\ ticker-days after false-discovery-rate control). The intensity of mark $k$ is
\begin{equation}
\lambda_k(t)=\mu_k+\sum_j a_{kj}\,\beta\sum_{t^j_i<t}e^{-\beta(t-t^j_i)},
\label{eq:hawkes}
\end{equation}
where $a_{kj}$ is the branching ratio and $\beta$ the decay rate. The log-likelihood is maximised exactly (Numba recursion for the univariate case, exact pairwise likelihood in PyTorch \citep{paszke2019} for the multivariate case with a shared $\beta$ and an $L_1$ penalty on $a$). The estimator recovers the parameters of data simulated by thinning \citep{ogata1981}. The pooled stream rejects a Poisson null by a parametric bootstrap ($p=\hkBootP$) and the per-ticker AIC prefers the Hawkes model for \hkPrefer\ of \NStocks\ stocks, with a median branching ratio of \hkMedAlpha. Event times lie on a 5-min grid, so $\beta$ is capped at \hkBetaCap~day$^{-1}$ (reached by \hkAtBound\ fits). For the network the intensity at the close of each day is computed with parameters fitted on the training window of the fold only; the filter is causal, which a truncation test confirms. Out of sample, the multivariate fit beats a Poisson model on \hkMvFolds\ of \NFolds\ folds, the univariate fit on \hkUniFolds, with a maximum spectral radius of \hkRhoMax.

\subsection{Covariance parametrisation and the PINC model}
\label{sec:pinc}
Every neural covariance has the form
\begin{equation}
\hat\Sigma_{t+1}=D\,\Phi\,D,\quad D=\mathrm{diag}\big(e^{h_i/2}\big),\quad
\Phi=(1-\pi)\Phi^{\rm RMT}_t+\pi\,\Phi^{\rm cc}_t,
\label{eq:sigma}
\end{equation}
where $h_i$ is the predicted log-variance of asset $i$, $\pi\in(0,1)$ is a predicted shrinkage level and $\Phi^{\rm cc}_t=\bar\rho_t\mathbf{1}\mathbf{1}^\top+(1-\bar\rho_t)I$ is the constant-correlation matrix. Both correlation matrices are symmetric positive semi-definite with unit diagonal, so $\hat\Sigma\succ0$ for any network output; the smallest eigenvalue over all forecasts is \minEig. The training loss is the matrix QLIKE \citep{patton2011},
\begin{equation}
L(\hat\Sigma,\RC)=\tfrac1N\big[\ln\det\hat\Sigma+\tr(\hat\Sigma^{-1}\RC)\big],
\label{eq:qlike}
\end{equation}
evaluated through a Cholesky factorisation. The network therefore learns volatilities and an amount of shrinkage around a data-driven correlation, a deliberately low-dimensional output given the \trainDays\ training days per fold.

Inputs are nine per-asset channels (HAR-type volatility lags, the negative part of the return, three jump measures, the Hawkes intensity and the Tsallis index) and seven global channels (market volatility, a regime flag, four spectral summaries of the correlation matrix and the mean Hawkes intensity), standardised with training-window statistics only. The \emph{PlainNN} is a shared multilayer perceptron over assets with a per-asset level, a linear skip connection (so that linear HAR-type predictors are nested) and a two-layer non-linear block. The \emph{LSTM} reads the last \seqLen\ days with the same output heads. The number of parameters at 32 hidden units is \parPlain\ (PlainNN), \parPINC\ (PINC) and \parLSTM\ (LSTM).

\emph{PINC} is exactly the PlainNN architecture trained with an additional regulariser inspired by the Fokker--Planck equation of an Ornstein--Uhlenbeck diffusion for log-volatility, $dh=-\kappa(h-\theta)dt+\xi\,dW$, whose stationary density is Gaussian and whose discrete form is an AR(1) with coefficient $\phi=e^{-\kappa}$. On a block of consecutive predictions $h_t$ it penalises the departure from that diffusion through a drift term $A=\overline{(h_{t+1}-\theta-\phi(h_t-\theta))^2}$ and a stationary-density term $B=\overline{(\mathrm{Var}_t(h)-s^2/(1-\phi^2))^2}$, where $s^2$ is the innovation variance of the block, with learnable $\phi$ (shared) and $\theta_i$. The total loss is
\begin{equation}
\mathcal{L}=\overline{L}+\alpha\,\tfrac12\big(A/s_A+B/s_B\big),
\label{eq:loss}
\end{equation}
with fixed data-based scales $s_A$, the mean training variance of log-volatility, and $s_B=s_A^2$. The weight $\alpha\in\{0,0.1,1,10\}$ is selected by the validation QLIKE of the data term only, and $\alpha=0$ is always in the grid, so the model may choose no physics. This is a regularisation \emph{inspired} by the Fokker--Planck equation; it does not minimise a PDE residual and the model is not a neural stochastic differential equation. Because the differences enter only between consecutive days, mini-batches are contiguous time blocks, never shuffled samples; a unit test shows that the term is near zero for a noise-free OU path, equals the innovation variance for a simulated AR(1) and is several times larger for the same values in shuffled order. With $\alpha=0$ PINC is bit-identical to the PlainNN.

Networks are trained with AdamW \citep{loshchilov2019}, gradient clipping at \gradClip, early stopping on validation \qlike\ (patience \patience, at most \maxEpochs\ epochs) and restoration of the best checkpoint. Hyper-parameters (hidden size, dropout, learning rate, weight decay, scheduler and, for the LSTM, the sequence length) are tuned with a seeded TPE search \citep{akiba2019} with the same budget of \hpoTrials\ trials per model and fold; PINC reuses the PlainNN hyper-parameters, so that $\alpha$ is the only difference. All models run on the CPU because for these small networks the CPU epoch time is \cpuSpeedMin--\cpuSpeedMax\ times shorter than on the Apple MPS back end (measured, Supplement).

\subsection{Benchmarks}
\label{sec:benchmarks}
The sample, Ledoit--Wolf and RMT estimators are recomputed from past windows of \corrWin\ days. \emph{HAR-RV} \citep{corsi2009} is an OLS regression of $\log RV_{t+1}$ on the daily, weekly and monthly means of $\log RV$, re-estimated on the training window of every fold, with a log-normal bias correction and the same RMT correlation base as the networks, so that the comparison with the networks isolates the volatility block. The extreme-value benchmark fits a generalised Pareto distribution by maximum likelihood to the exceedances over a threshold chosen on validation data \citep{mcneilfrey2000}; it scales the RMT correlation by the 5\,\% expected shortfall of each asset; fits with too few exceedances fall back to the historical quantile and are counted (\gpdFallback\ of \gpdFits). Rolling estimators are calibrated to the scale of the realised covariance by a single scalar per fold fitted on validation days.

\subsection{Evaluation protocol}
\label{sec:protocol}
The sample is split in \NFolds\ outer folds with a training window of \trainDays, a validation window of \valDays\ and a test window of \testWin\ days; test windows are contiguous and give \NTestDays\ out-of-sample decision days from \testStart\ to \testEnd. Every model with estimated parameters (HAR, LSTM, PlainNN, PINC, Hawkes) is re-estimated at every fold boundary. Scalers, Hawkes parameters, HAR coefficients, the EVT threshold, $\alpha$ and early stopping use training and validation data only. As a leakage test, all data from the first test day on were replaced by absurd values; training of the first fold remained bit-identical. Neural models are trained with \NSeeds\ seeds and the results are averaged.

\subsection{Portfolios}
\label{sec:portfolios}
All allocators are long-only, fully invested, with a single-name cap of \capW, a turnover limit $\|w-w^{\rm drift}\|_1\le\turnLim$ with respect to the drifted previous weights, and a small ridge towards them. Minimum-variance portfolios solve a quadratic programme for each covariance forecast. Equal-risk-contribution (ERC) portfolios \citep{maillard2010} use the RMT covariance, and $1/N$ is held either rebalanced or buy-and-hold. The EVaR portfolio minimises the entropic VaR \citep{ahmadijavid2012}, $\inf_{z>0}z\ln(\mathbb{E}[e^{L/z}]/\alpha)$, as an exponential-cone programme on the last \evarScen\ historical return scenarios; controlled tests confirm that it equals a one-dimensional minimisation, that $\mathrm{EVaR}\ge\mathrm{ES}\ge\mathrm{VaR}$ and that it recovers $\sigma\sqrt{-2\ln\alpha}$ for Gaussian losses. The problems are solved with CVXPY \citep{diamondboyd2016}; any solver failure would keep the previous weights and be counted, and there were none in \nOpt\ optimisations. Transaction costs of \costBps\,bps per unit traded are subtracted \emph{after} the optimisation, so decisions do not depend on the cost level. The risk-free rate is set to zero.

\subsection{Metrics and tests}
\label{sec:tests}
Forecasts are scored on test days by the QLIKE of Eq.~(\ref{eq:qlike}), the Frobenius error $\|\hat\Sigma-\RC\|_F$ and the Gaussian log-likelihood of close-to-close returns. Differences of daily losses between PINC and each comparator are tested with a Diebold--Mariano test \citep{dieboldmariano1995} using a Newey--West variance \citep{neweywest1987} and the small-sample correction of \citet{harvey1997}. Portfolio differences in Sharpe ratio, volatility and maximum drawdown use the paired stationary bootstrap \citep{politisromano1994} (mean block \bootBlock\ days, \nBoot\ resamples, one-sided $p$-values from the re-centred distribution). Each family (one metric against all comparators) is adjusted by the Holm procedure \citep{holm1979}; Benjamini--Hochberg $q$-values \citep{benjamini1995} are reported in the Supplement. Only adjusted values are used for statements of superiority. VaR and ES forecasts for day $t+1$ use only past losses under the weights about to be held (historical, extreme-value and Gaussian-$\hat\Sigma$ families) and are tested with the Kupiec unconditional-coverage test \citep{kupiec1995} and the Christoffersen conditional-coverage test \citep{christoffersen1998}; ES is scored with the Fissler--Ziegel loss \citep{fisslerziegel2016}. EVaR dominates VaR and ES, so a correct EVaR should be conservative and its violation rate is compared with $\alpha$ rather than tested for equality.

\section{Results}
\label{sec:results}

\subsection{Covariance forecasts}
\label{sec:res_forecast}
\begin{figure*}[t]
\centering
\includegraphics[width=0.9\textwidth]{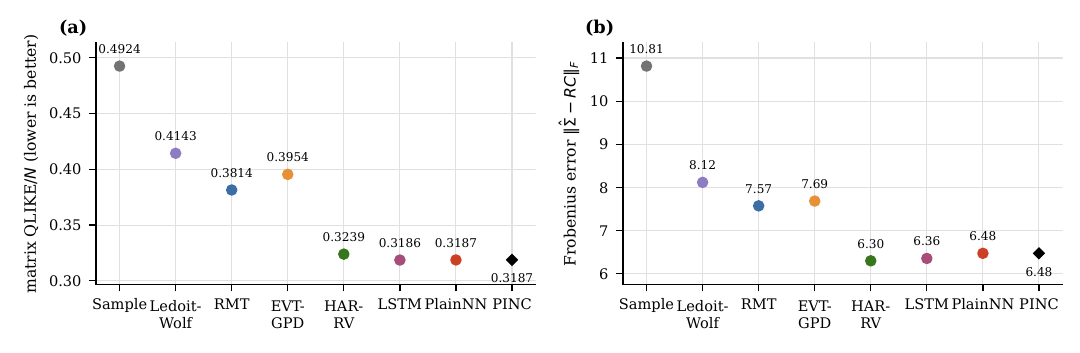}
\caption{Out-of-sample covariance accuracy. (a) Matrix QLIKE$/N$ and (b) Frobenius error; markers are means over days (and seeds) and bars the standard deviation across seeds (barely visible). Lower is better. PINC and PlainNN coincide.}
\label{fig:forecast}
\end{figure*}

Table~\ref{tab:forecast} and Fig.~\ref{fig:forecast} report out-of-sample accuracy. The ordering by \qlike\ is Sample ($\qlSample$), Ledoit--Wolf ($\qlLW$), EVT-GPD ($\qlEVT$), RMT ($\qlRMT$), HAR-RV ($\qlHAR$), PlainNN and PINC (both $\qlPINC$) and the LSTM ($\qlLSTM$, the lowest mean). The neural forecasts are significantly better than the four static estimators: Table~\ref{tab:dm} gives \qlike\ differences from $\dSample$ (Sample) to $\dEVT$ (EVT-GPD) with Holm-adjusted $p<0.001$. Against HAR-RV the \qlike\ difference is $\dHAR$ ($p_{\rm Holm}=\pHAR$) and against the LSTM $\dLSTM$ ($p_{\rm Holm}=\pLSTM$), so neither is significant. On the Frobenius error HAR-RV and the LSTM are \emph{better} than PINC (two-sided $p=\pFrobHARtwo$ and $p\pFrobLSTMtwo$ before adjustment), which shows that the small \qlike\ advantage over HAR-RV does not carry over to every loss. The seed-to-seed standard deviation of the neural \qlike\ is at most \qlsdPINC, much smaller than the differences with static estimators and also smaller than the difference with HAR-RV ($\dHAR$).

\begin{table}[t]
\centering
\small
\input{tab_forecast.tex}
\caption{Out-of-sample covariance forecasts (\NTestDays\ days, \NFolds\ folds). Mean of the daily losses, with the standard deviation across seeds for neural models. PINC and PlainNN have identical forecasts because $\alpha^*=0$ in every fold. Bold: lowest mean QLIKE.}
\label{tab:forecast}
\end{table}

\begin{table}[t]
\centering
\small
\input{tab_dm.tex}
\caption{Diebold--Mariano tests of PINC against each comparator (negative $\Delta$: PINC has the lower loss). $p_{\rm Holm}$: one-sided $p$-value of ``PINC better'' adjusted over the family of comparators.}
\label{tab:dm}
\end{table}

\subsection{The physics-inspired regulariser and the ablations}
\label{sec:res_ablation}
Validation selected $\alpha^*=0$ in all \NFolds\ folds: the mean validation \qlike\ rises monotonically with $\alpha$, from \valAlphaZero\ at $\alpha=0$ to \valAlphaTen\ at $\alpha=10$ (Fig.~\ref{fig:ablation}b; fold-wise values in the Supplement). The regulariser was therefore rejected before the test window was consulted, PINC coincides with the PlainNN, and no test-window comparison between them exists. For reporting only, the test \qlike\ for fixed positive weights is \qlAlphaA, \qlAlphaB\ and \qlAlphaC\ for $\alpha=0.1,1,10$ against \qlPlain\ at $\alpha=0$, always worse.

The ablations retrain the network in every fold with the same seeds and hyper-parameters (Fig.~\ref{fig:ablation}a). Removing the RMT correlation base and spectral inputs changes the \qlike\ by $\dabNoRMT$ ($p_{\rm Holm}=\pabNoRMT$), removing the Tsallis index by $\dabNoTsallis$ ($p_{\rm Holm}=\pabNoTsallis$) and removing the Hawkes intensity by $\dabNoHawkes$ ($p_{\rm Holm}=\pabNoHawkes$). The variants are nominally slightly better, but none of these differences is significant, so the inputs have no detectable effect on the forecast. Without RMT the Frobenius error is worse (\frNoRMT\ against \frPINC) and the minimum-variance volatility higher (\volNoRMT\,\% against \volFull\,\%), which illustrates that a lower \qlike\ does not imply a better portfolio.

\begin{figure*}[t]
\centering
\includegraphics[width=0.9\textwidth]{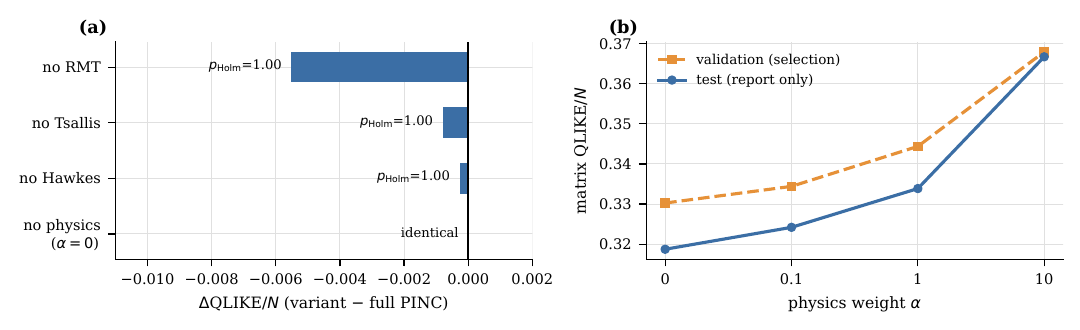}
\caption{(a) Change in test QLIKE$/N$ when an ingredient is removed (variant minus full PINC; negative: the variant is nominally better) with Holm-adjusted Diebold--Mariano $p$-values. (b) Validation (selection) and test (report only) QLIKE$/N$ as a function of the physics weight $\alpha$.}
\label{fig:ablation}
\end{figure*}

\subsection{Portfolios after costs}
\label{sec:res_portfolio}
\begin{figure*}[t]
\centering
\includegraphics[width=0.9\textwidth]{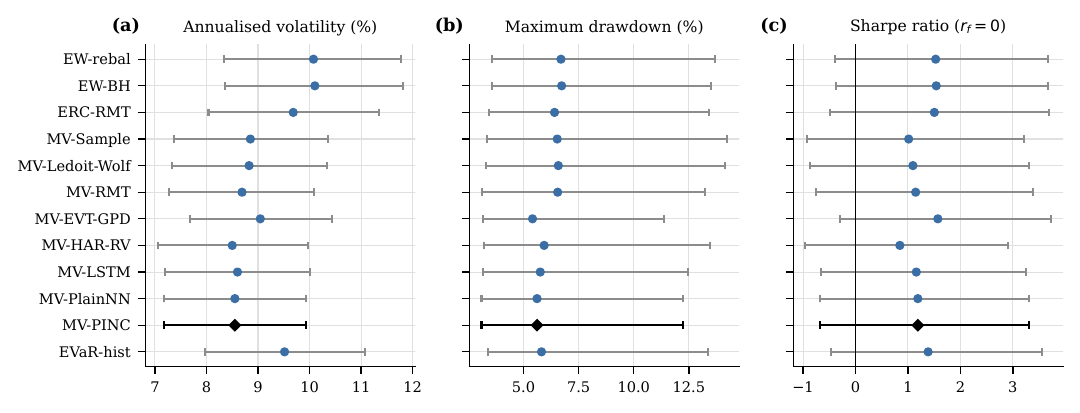}
\caption{Portfolio outcomes net of \costBps\,bps with 95\,\% stationary-bootstrap intervals (mean block \bootBlock\ days). Overlapping intervals show that most rankings are not established. MV-PINC is the black diamond.}
\label{fig:portfolio}
\end{figure*}

Table~\ref{tab:portfolio} and Fig.~\ref{fig:portfolio} give the net results at \costBps\,bps. The highest Sharpe ratio belongs to the minimum-variance portfolio on the EVT covariance (\shEVT), the lowest volatility to HAR-RV (\volHAR\,\%) and the smallest drawdown to EVT (\ddEVT\,\%). MV-PINC has a Sharpe ratio of \shPINC\ (95\,\% interval \shPINClo\ to \shPINChi), a volatility of \volPINC\,\% and a drawdown of \ddPINC\,\%. The paired bootstrap finds no significant Sharpe or drawdown difference between MV-PINC and any comparator after Holm adjustment; the closest case is HAR-RV, where PINC has a higher Sharpe by \dShHAR\ but $p_{\rm Holm}=\pHolmShHAR$. The only significant differences are in volatility: MV-PINC is less volatile than $1/N$ ($p_{\rm Holm}=\pHolmVolEW$), ERC, the EVT portfolio ($p_{\rm Holm}=\pHolmVolEVT$) and the historical EVaR portfolio, but not than the sample ($p_{\rm Holm}=\pHolmVolSample$), Ledoit--Wolf, HAR-RV or LSTM minimum-variance portfolios. The Sharpe intervals of all strategies overlap widely, as expected for \NTestDays\ days, so the ranking by Sharpe is not statistically established.

\begin{figure}[t]
\centering
\includegraphics[width=0.95\columnwidth]{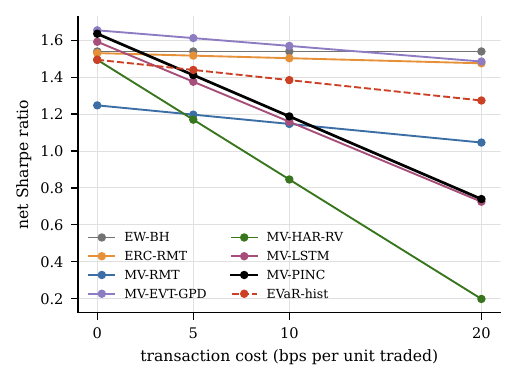}
\caption{Net Sharpe ratio as a function of the transaction cost. Strategies with high turnover lose most; PlainNN and PINC coincide.}
\label{fig:costs}
\end{figure}

Fig.~\ref{fig:costs} shows that the ranking by Sharpe depends strongly on costs. At zero cost the highest Sharpe is that of the EVT portfolio (\shEVTzero) and PINC reaches \shPINCzero, but the daily turnover of the neural and HAR portfolios (\toPINC\ and \toHAR\ against \toRMT\ for RMT and \toEW\ for $1/N$) reduces PINC to \shPINCtwenty\ and HAR-RV to \shHARtwenty\ at 20\,bps, where buy-and-hold $1/N$ (\shEWBHtwenty) is the best strategy. The portfolio weights of the neural models are close to each other (mean daily $L_1$ distance \lonePINCLSTM\ between PINC and LSTM against \lonePINCHAR\ for HAR-RV and \lonePINCRMT\ for RMT).

\begin{table*}[t]
\centering
\small
\input{tab_portfolio.tex}
\caption{Out-of-sample portfolios net of \costBps\,bps costs (\NTestDays\ days). Return and drawdown in \% (annualised for the return and volatility); 95\,\% interval for the Sharpe ratio from the stationary bootstrap; turnover is the mean daily $\ell_1$ trade; $N_{\rm eff}=1/\sum w^2$. Bold: best value in the column among all strategies. Neural strategies are averaged over \NSeeds\ seeds; PINC and PlainNN coincide.}
\label{tab:portfolio}
\end{table*}

\subsection{Portfolio-risk forecasts and tail risk}
\label{sec:res_tail}
The covariance forecast is used by minimum-variance investors through $w^\top\hat\Sigma w$. Evaluated on each strategy's own weights, the correlation between predicted and realised portfolio volatility is \corrPINC\ for PINC, \corrLSTM\ for the LSTM, \corrHAR\ for HAR-RV and \corrRMT\ for RMT, and the ratio of predicted to realised variance is \biasPINC\ for PINC and \biasRMT\ for RMT (close to 1 for PINC and the LSTM, below 1 for RMT, EVT-GPD and HAR-RV, and well above 1 for the sample covariance; Supplement).

Historical-simulation VaR is calibrated for most strategies but not all: at 1\,\% (expected violations \expOne\ in \NTestDays\ days) the Kupiec test rejects \kupOne\ of \NStrat\ strategies and at 5\,\% (expected \expFive) \kupFive. For PINC the numbers of violations are \violOnePINC\ and \violFivePINC, with Kupiec $p$-values of \kupPOnePINC\ and \kupPFivePINC. The power of the independence test at 1\,\% is very low with so few violations. The EVaR at 5\,\% of the backcast sample is conservative, as it should be, for all \NStrat\ strategies (violation rates \evarMin\ to \evarMax, all below the nominal 0.05). The EVaR-optimal portfolio differs from the minimum-variance portfolios on real data: the $L_1$ distance between their weights at the last decision date is \evarLoneLo--\evarLoneHi\ for three levels and the distance between the 1\,\% and 10\,\% EVaR weights is \evarLoneAlpha, so the weights depend on the tail level. This portfolio is computed from only \evarScen\ scenarios and its historical VaR at 1\,\% is rejected by the Kupiec test (\violEVaRhistOne\ violations, $p\kupEVaRhistOne$).

\subsection{Robustness}
\label{sec:res_robust}
Scoring the same forecasts against realised covariances built from 10-, 15- and 30-min returns changes the leader: the LSTM has the lowest \qlike\ at 5, 10 and 15\,min and HAR-RV at 30\,min, while PINC ranks \rankPINCfive\ at 5\,min and \rankPINCthirty\ at 30\,min among eight models (Supplement). Across \nCapSettings\ combinations of weight cap and turnover limit, PINC has a higher Sharpe ratio than the RMT portfolio in \capShRMT\ and than the PlainNN in \capShPlain, the latter because the two coincide. The qualitative conclusions are therefore not sensitive to these choices, but the ordering among the leading models is.

\section{Discussion}
\label{sec:discussion}

\emph{Why was the regulariser rejected?} Eq.~(\ref{eq:loss}) pulls the sequence of predicted log-variances towards an AR(1)/Ornstein--Uhlenbeck path with a Gaussian stationary law. The network already receives HAR-type lags of realised volatility and a linear skip connection, so its forecasts follow the persistence of volatility closely; any further constraint is an additional bias, and validation \qlike\ rose monotonically with $\alpha$. With \trainDays\ training days per fold and a one-day horizon there are few independent observations from which a stationary-density term could help. We make no claim that a physics-based prior cannot help in other settings (longer samples, other markets, a genuine PDE residual, other horizons).

\emph{What can and cannot be concluded.} The evidence supports three statements: the network improves on static covariance estimators; it is not distinguishable from HAR-RV or an LSTM in \qlike\ and is worse on the Frobenius error; and removing the RMT, Tsallis or Hawkes inputs has no detectable effect. It does not support the statement that these tools are useless, because the test has limited power: \NTestDays\ days cover one year in one market regime, and a difference of the size separating PINC from HAR-RV ($\dHAR$) would require a longer sample to be established. The Hawkes input is also weakened by the 5-min grid of the event times and by the small number of events per stock (\hkEvMin--\hkEvMax\ per stock in the sample), and the Tsallis index is a re-parametrisation of the Student-$t$ degrees of freedom. The portfolio comparison is limited by the same sample size, the Sharpe intervals span several units, and the ranking changes with costs, so we do not recommend any strategy on these data.

\emph{Relation to physics-informed networks.} Physics-informed networks \citep{raissi2019} enforce a differential equation through its residual at collocation points. The present regulariser is a moment-based consistency condition and the model has no density network or PDE residual, so the term ``physics-informed'' would be misleading; we use ``physics-inspired''. That such a weak constraint is rejected by validation does not speak for or against full PDE-constrained models.

\emph{Limitations.} (i) Hawkes event times lie on a 5-min grid and $\beta$ is capped at \hkBetaCap~day$^{-1}$; the multivariate fit shares one $\beta$ and uses an $L_1$ penalty. (ii) The regulariser is not a PDE residual. (iii) The risk-free rate is zero. (iv) Only \NTestDays\ out-of-sample days in a single period and universe are used. (v) The hyper-parameter budget is \hpoTrials\ trials per model and fold, and PINC receives extra trainings for $\alpha$. (vi) The EVaR scenario programme is solved with few scenarios (\evarScen); at several thousand scenarios the solver failed in our tests. (vii) The independence test of VaR is not informative at 1\,\%. (viii) Intraday realised covariances are sensitive to microstructure noise, which we mitigate by sub-sampling. (ix) The multiple-comparison correction is applied within each family of comparisons, not across all analyses in the paper.

\section{Summary and conclusions}
\label{sec:conclusions}

In a nested walk-forward experiment with \NFolds\ folds and \NTestDays\ out-of-sample days on the \NStocks\ DJIA stocks, we found that (Q1) removing the RMT, Tsallis or Hawkes inputs has no detectable effect on the covariance forecast; (Q2) the Fokker--Planck-inspired regulariser is rejected by validation in every fold, so PINC equals the plain network; (Q3) the neural forecast is significantly better than sample, Ledoit--Wolf, RMT and EVT covariances but not significantly different from HAR-RV or an LSTM in \qlike; (Q4) no Sharpe or drawdown difference between the PINC minimum-variance portfolio and the benchmarks is significant after correction, and the ranking depends on costs; (Q5) historical VaR is rejected by the Kupiec test for \kupOne\ of \NStrat\ strategies at 1\,\%, the EVaR of the backcast sample is conservative, and EVaR weights depend on the tail level; (Q6) the best model depends on the target sampling frequency.

The practical message is that, for this data set and horizon, a HAR-type benchmark captures what the econophysics-inspired inputs and regulariser offered, and that claims of the opposite should be accompanied by an equally strong benchmark, a same-capacity network without the physics term and a protocol that excludes temporal leakage. Natural extensions are longer and multi-market samples, longer horizons, a genuine PDE-constrained density model, and an event-time Hawkes estimator without the 5-min grid.

\section*{Data and code availability}

The intraday dataset used in this study consists of third-party financial data and is subject to the applicable data-provider licensing conditions. Consequently, the raw dataset is not redistributed with this article.
The source code, computational notebooks, experimental configurations and derived results supporting the findings of this study are available from the corresponding author upon reasonable request, subject to applicable data licensing restrictions.
The corresponding author may be contacted to request further information concerning the implementation, experimental procedures and computational materials required to reproduce the analyses. All numerical results and tables reported in this paper are derived from the corresponding experimental outputs.

\section*{Computing environment}
Experiments ran on \cpuName\ (\ramGB\,GB RAM), Python \pyVer, PyTorch \torchVer

\section*{Declaration of competing interest}
The author declares no competing interests.

\bibliographystyle{elsarticle-harv}
\bibliography{refs}

\end{document}


\maketitle

\noindent This document collects the validation outputs, tables and figures that support the main text. Every number is produced by the same script that produces the main text and is read from the CSV files of the run (\NFolds\ folds, \NTestDays\ test days, \NStocks\ stocks). Sections S1--S6 follow the order in which they are cited in the manuscript.

\section{S1. Tests on synthetic data and temporal-separation checks}
Before any real data are used, the code is checked on simulated data: the Hawkes maximum-likelihood estimator is run on data simulated by Ogata thinning, the temporal-separation tests include a negative control that must fail, and the positive-definiteness of the covariance output is verified. The output of those cells is reproduced below (non-ASCII characters have been replaced).

{\footnotesize
\begin{Verbatim}
### Cell 10 - temporal separation and look-ahead tests
folds available: 11 | used: 11 (FULL experiment)
 negative control: same-timestamp features/target correctly rejected
 temporal-separation tests passed | test days used: 231

### Cell 15 - Hawkes MLE recovered on simulated Ogata-thinning data
Poisson truth (alpha=0): mean estimated alpha = 0.087, 90th percentile = 0.249  (finite-sample
    positive bias is expected since alpha>=0)
per-ticker: median alpha = 0.131, share with alpha>0.01 = 1.00, tickers with dAIC<-2 (Hawkes
    preferred) = 28/29 | log: {'mv_rescaled_for_stability': 0, 'uni_hessian_not_pd': 2,
    'uni_at_alpha_bound': 0, 'mv_not_converged': 0}
pooled event stream: LR = 3977.74; parametric-bootstrap p-value (B=500) = 0.0020; null LR 95th
    percentile = 6.89
held-out test windows (params fitted on train only; positive = better than Poisson):
    multivariate beats Poisson in 7/11 folds (mean dLL +3.89), univariate in 9/11 (mean +6.16) |
    max rho = 0.320 | log: {'mv_rescaled_for_stability': 0, 'uni_hessian_not_pd': 2,
    'uni_at_alpha_bound': 0, 'mv_not_converged': 0}
causality test passed: intensity up to day 300 is unchanged when all later events are removed.

### Cell 20 - HAR-RV regression recovered on a simulated HAR process
simulated HAR: true beta = [0.35 0.3  0.3 ] | OLS beta = [0.359 0.324 0.267] | sigma^2 = 0.0893
    (true 0.09)

### Cell 21 - GPD-POT VaR/ES against the exact Student-t4 values
EW portfolio, first 252 days: GPD xi = -0.116, sigma = 0.0050 | 5% VaR = 1.00%, ES = 1.40% |
    historical 5% VaR = 0.91%

### Cell 23 - checks of the Fokker-Planck-inspired terms
exact noise-free OU path:  A = 5.98e-15 (must be ~ 0)
stationary AR(1) with true phi=0.880: A = 0.0901 (~ innovation var 0.09), B = 0.0005 (~ 0; stat.
    var 0.399)
same values in SHUFFLED order: A = 0.7159  (>> 0.0901: the term is only meaningful for time-
    ordered blocks)
gradients w.r.t. predictions are finite; PINC parameters: 1742 vs PlainNN: 1712 (difference =
    phi and theta_i only; unused when alpha=0)

### Cell 29 - leakage test (training with test-window data destroyed)
leakage test passed: with all data from 2014-02-06 on destroyed, fold-0 training is bit-
    identical (val loss 0.543730).

### Cell 32 - EVaR cone programme: controlled tests
test 1-3 passed: cone = 1-D minimiser (< 1e-4), EVaR >= ES >= VaR, increasing as alpha shrinks,
    matches sigmasqrt(-2 ln alpha) for Gaussian losses.
Gaussian 5 assets (S=3000 scenarios): EVaR(5%) weights [0.325 0.046 0.146 0.234 0.249] | min-
    variance [0.323 0.054 0.139 0.226 0.258] | L1 distance 0.034
(scenario-EVaR weights carry sampling noise that shrinks with S: L1 distance 0.19 / 0.12 / 0.07
    for S = 1500 / 2000 / 3000 in a check run; with the 126 scenarios used in the backtest EVaR
    weights are noisier than min-variance weights - a property to read in Cell 36)
asset variances: A 0.99, B 0.71 (B is the *lower*-variance asset but carries a 3 % chance of a
    -4 loss)

### Cell 39 - size and power of the Diebold-Mariano implementation
DM size under H0 (n=120, 1000 sims): rejection rate 0.056 (nominal 0.05) | power at mean
    -0.4sigma: 0.99
\end{Verbatim}}

\section{S2. Data and microstructure}
\begin{figure}[p]
\centering\includegraphics[width=\textwidth]{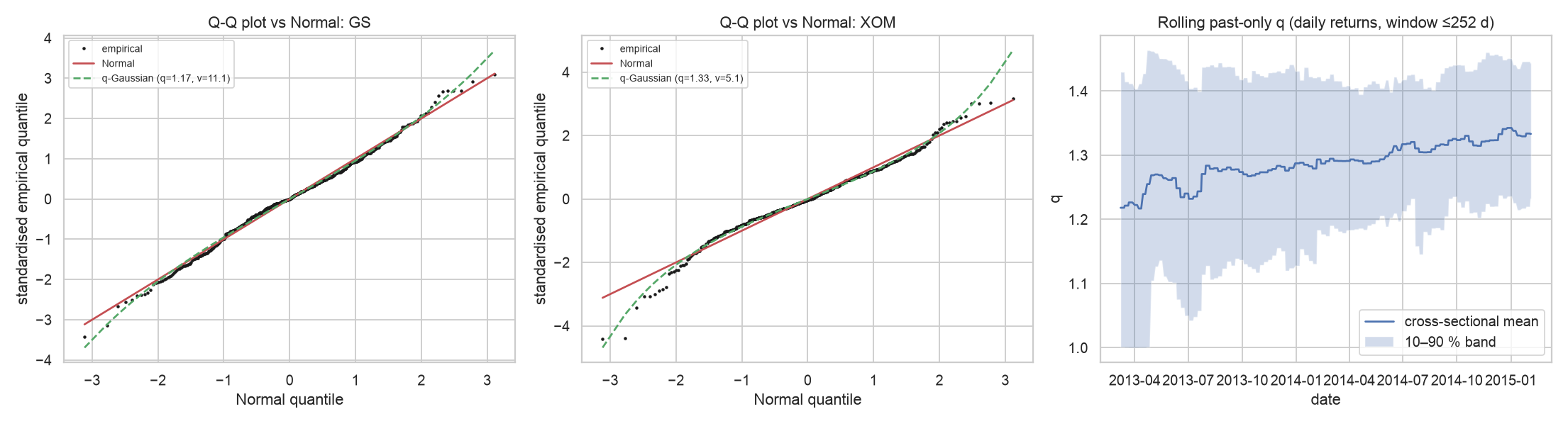}
\caption{Tail behaviour of the standardised 5-min returns and $q$-Gaussian fits.}\label{fig:s_tails}
\end{figure}

\begin{figure}[p]
\centering\includegraphics[width=\textwidth]{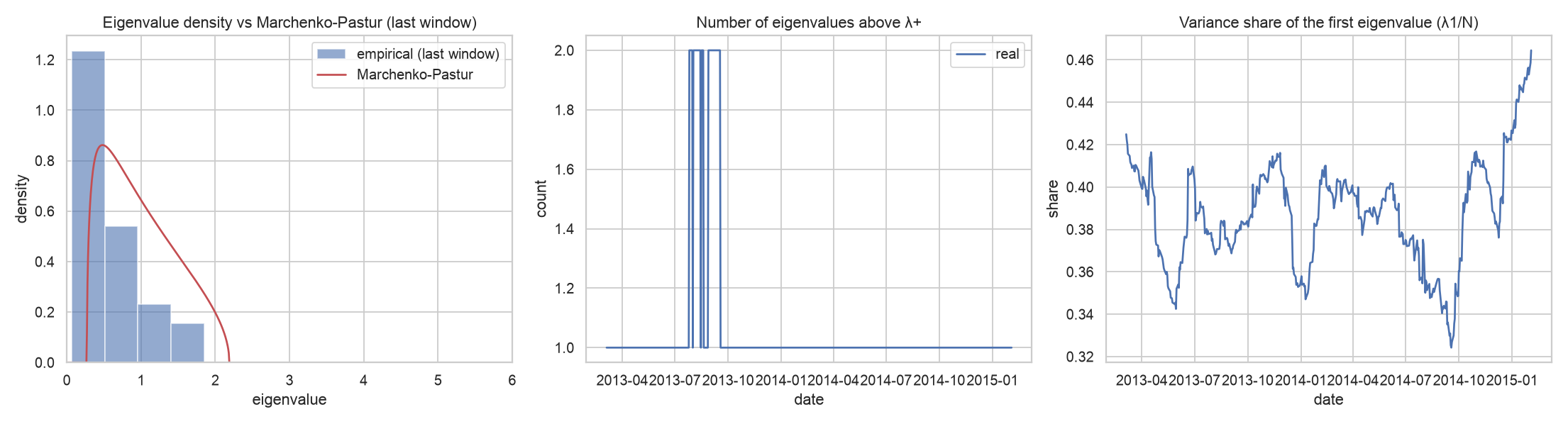}
\caption{Random-matrix diagnostics of the correlation matrix.}\label{fig:s_rmt}
\end{figure}

\begin{figure}[p]
\centering\includegraphics[width=\textwidth]{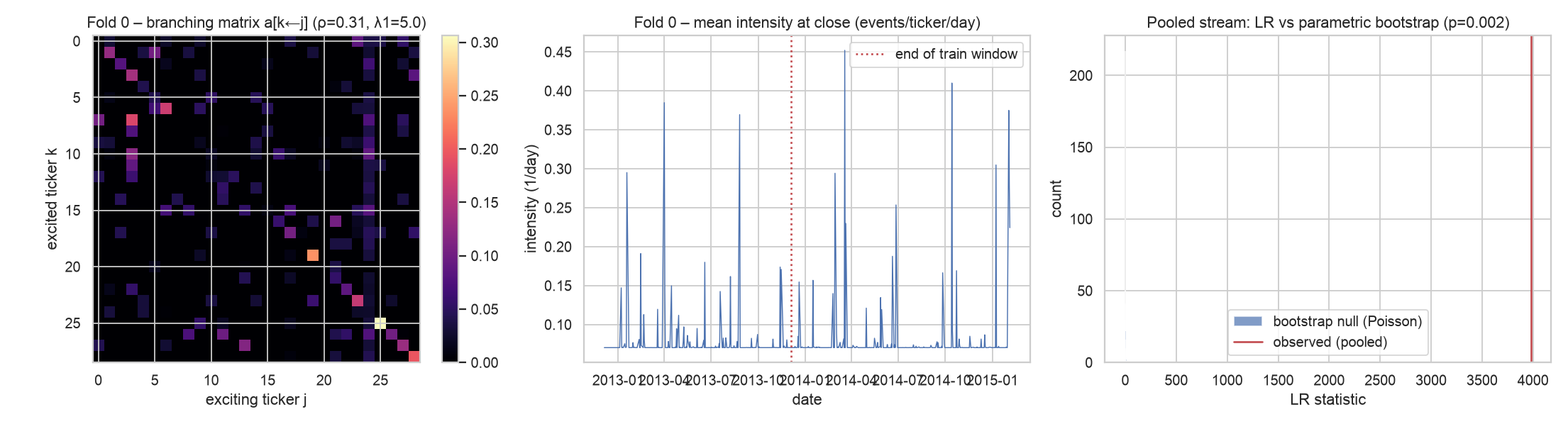}
\caption{Hawkes self-excitation of extreme events: per-ticker branching ratios and out-of-sample comparison with a Poisson model.}\label{fig:s_hawkes}
\end{figure}

Fig.~\ref{fig:s_tails} shows the tail fits of the 5-min returns, Fig.~\ref{fig:s_rmt} the RMT diagnostics (spectrum of the correlation matrix against the Marchenko--Pastur bulk, with a permuted control) and Fig.~\ref{fig:s_hawkes} the Hawkes fits. The pooled stream of \hkNEvents\ events rejects a Poisson null by a parametric bootstrap ($p=\hkBootP$); the per-ticker AIC prefers the Hawkes model for \hkPrefer\ of \NStocks\ stocks. The number of events per stock lies between \hkEvMin\ and \hkEvMax, which limits the information content of this input.


\section{S3. Hyper-parameters, physics weight and computing devices}
Table~\ref{tab:s_alpha} gives the validation \qlike\ for each value of the physics weight $\alpha$ in each fold; $\alpha=0$ is selected in every fold, so PINC and PlainNN produce identical forecasts. Table~\ref{tab:s_device} compares training time per epoch on CPU and on the Apple MPS back-end; all neural models in the paper run on CPU.

\begin{table}[h]\centering\caption{Mean validation \qlike$/N$ by physics weight $\alpha$ and fold.}\label{tab:s_alpha}
\sizedtab{tabS_alpha.tex}\end{table}
\begin{table}[h]\centering\caption{Training time per epoch by model and device, and validation \qlike.}\label{tab:s_device}
\sizedtab{tabS_device.tex}\end{table}

\section{S4. Full VaR and expected-shortfall backtests}
Table~\ref{tab:s_var} reports violations, Kupiec test and Christoffersen conditional-coverage test at the 1\,\% and 5\,\% levels for each of the 12 strategies, together with the coverage of the EVaR bound at 5\,\%.

\begin{table}[h]\centering\caption{VaR backtests by strategy (daily portfolio returns, \NTestDays\ days).}\label{tab:s_var}
\sizedtab{tabS_var.tex}\end{table}

\section{S5. Robustness}
\subsection*{S5.1 Realised-covariance sampling frequency}
\begin{table}[h]\centering\caption{Mean matrix \qlike$/N$ when the realised covariance is built from 5-, 10-, 15- and 30-min returns.}\label{tab:s_freq}
\sizedtab{tabS_freq.tex}\end{table}
\subsection*{S5.2 Transaction costs}
\begin{table}[h]\centering\caption{Net Sharpe ratio of each strategy for different proportional costs.}\label{tab:s_costs}
\sizedtab{tabS_costs.tex}\end{table}
\subsection*{S5.3 Predicted versus realised risk of each strategy's own portfolio}
\begin{table}[h]\centering\caption{Annualised realised and predicted volatility (\%) of the minimum-variance portfolio of each covariance model; bias is the ratio of predicted to realised variance; Corr.\ is the correlation between predicted and realised volatility.}\label{tab:s_own}
\sizedtab{tabS_ownrisk.tex}\end{table}

\begin{figure}[p]
\centering\includegraphics[width=\textwidth]{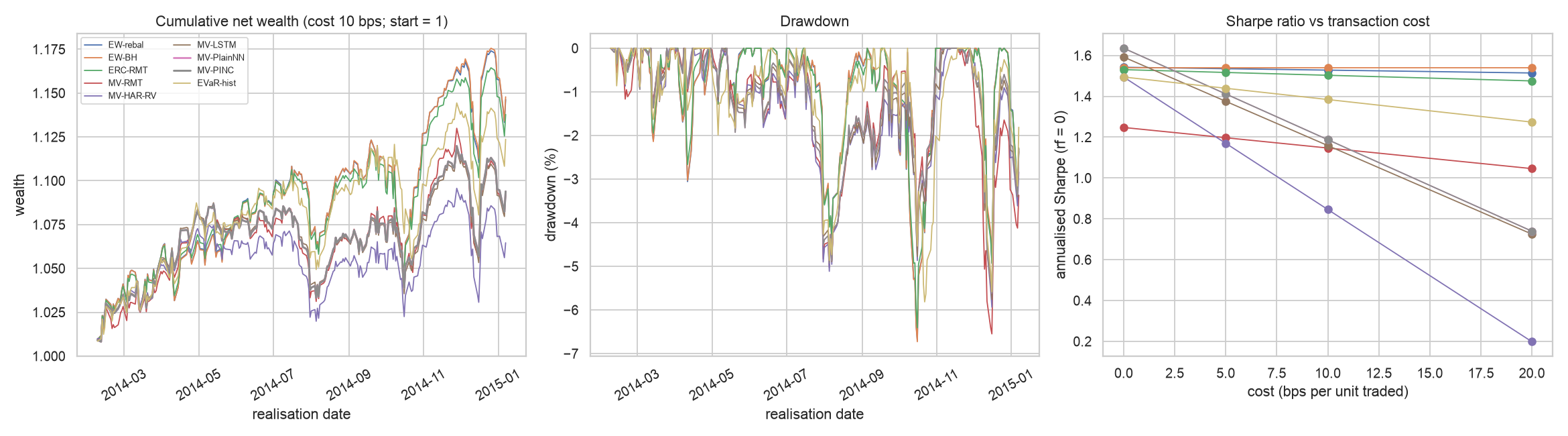}
\caption{Backtest: cumulative net returns and drawdowns of the strategies.}\label{fig:s_backtest}
\end{figure}
\begin{figure}[p]
\centering\includegraphics[width=\textwidth]{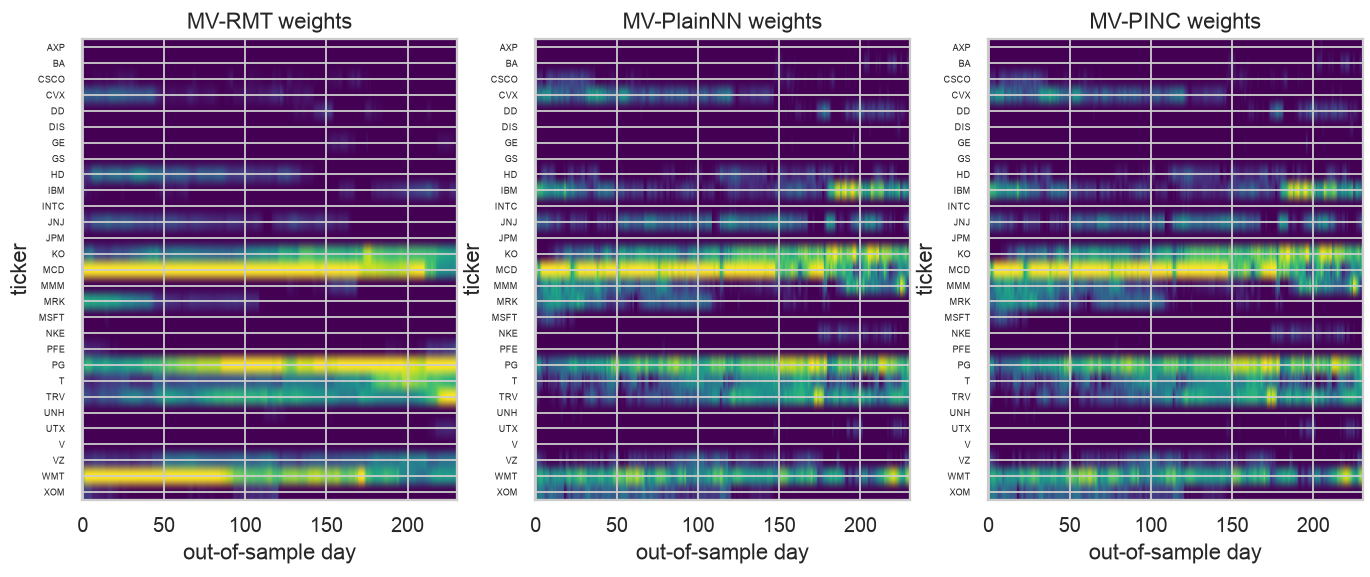}
\caption{Portfolio weights through time for three of the strategies.}\label{fig:s_weights}
\end{figure}

\section{S6. Audit and complete test tables}
\subsection*{S6.1 Self-audit of the pipeline}
\begin{table}[h]\centering\caption{Requirements verified automatically at the end of the run.}\label{tab:s_audit}
{\small\input{tabS_audit.tex}}\end{table}
\subsection*{S6.2 Diebold--Mariano tests}
\begin{table}[h]\centering\caption{PINC against each comparator: mean loss difference $\Delta$ (PINC minus comparator), Diebold--Mariano statistic, one-sided $p$ ($p_1$) and Holm-adjusted $p$ within each family.}\label{tab:s_dm}
\sizedtab{tabS_dm.tex}\end{table}
\subsection*{S6.3 Ablations}
\begin{table}[h]\centering\caption{Ablation variants against the full PINC (same seeds and hyper-parameters). $p_1$ is the one-sided Diebold--Mariano $p$ for ``full model better''.}\label{tab:s_abl}
\sizedtab{tabS_ablation.tex}\end{table}

%% file: numbers.tex
\newcommand{\NDays}{545}
\newcommand{\NFolds}{11}
\newcommand{\NHalfDays}{7}
\newcommand{\NMinutes}{211{,}290}
\newcommand{\NSeeds}{3}
\newcommand{\NStocks}{29}
\newcommand{\NStrat}{12}
\newcommand{\NTestDays}{231}
\newcommand{\PctImputed}{0.03}
\newcommand{\PctLost}{0.96}
\newcommand{\aicQNormal}{29}

\newcommand{\biasPINC}{1.03}
\newcommand{\biasRMT}{0.97}

\newcommand{\bnsJumpDays}{29}
\newcommand{\bnsTickerDays}{15{,}805}
\newcommand{\bootBlock}{5}
\newcommand{\capShPlain}{0}
\newcommand{\capShRMT}{6}
\newcommand{\capW}{0.20}

\newcommand{\corrHAR}{0.55}
\newcommand{\corrLSTM}{0.63}
\newcommand{\corrPINC}{0.62}
\newcommand{\corrRMT}{0.38}
\newcommand{\corrWin}{126}
\newcommand{\costBps}{10}
\newcommand{\cpuName}{Apple M3 Pro}
\newcommand{\cpuSpeedMax}{10.9}
\newcommand{\cpuSpeedMin}{1.5}
\newcommand{\dEVT}{\ensuremath{-}0.0766}
\newcommand{\dHAR}{\ensuremath{-}0.0052}
\newcommand{\dLSTM}{0.0001}

\newcommand{\dSample}{\ensuremath{-}0.1737}
\newcommand{\dShHAR}{0.34}
\newcommand{\dabNoHawkes}{\ensuremath{-}0.0003}
\newcommand{\dabNoRMT}{\ensuremath{-}0.0055}
\newcommand{\dabNoTsallis}{\ensuremath{-}0.0008}

\newcommand{\ddEVT}{5.41}

\newcommand{\ddPINC}{5.61}

\newcommand{\evarLoneAlpha}{0.274}
\newcommand{\evarLoneHi}{0.420}
\newcommand{\evarLoneLo}{0.376}
\newcommand{\evarMax}{0.039}
\newcommand{\evarMin}{0.009}
\newcommand{\evarScen}{126}
\newcommand{\expFive}{11.6}

\newcommand{\expOne}{2.3}

\newcommand{\frNoRMT}{6.635}
\newcommand{\frPINC}{6.475}

\newcommand{\gpdFallback}{35}
\newcommand{\gpdFits}{12{,}881}
\newcommand{\gradClip}{1.0}
\newcommand{\hkAtBound}{2}
\newcommand{\hkBetaCap}{78}
\newcommand{\hkBootP}{0.002}
\newcommand{\hkEvMax}{76}
\newcommand{\hkEvMin}{32}

\newcommand{\hkMedAlpha}{0.131}
\newcommand{\hkMvFolds}{7}
\newcommand{\hkNEvents}{1{,}371}

\newcommand{\hkPrefer}{28}
\newcommand{\hkRhoMax}{0.320}
\newcommand{\hkUniFolds}{9}

\newcommand{\hpoTrials}{12}
\newcommand{\kupEVaRhistOne}{\ensuremath{<}0.001}
\newcommand{\kupFive}{2}
\newcommand{\kupOne}{4}
\newcommand{\kupPFivePINC}{0.473}
\newcommand{\kupPOnePINC}{0.123}

\newcommand{\lonePINCHAR}{0.31}
\newcommand{\lonePINCLSTM}{0.11}
\newcommand{\lonePINCRMT}{0.53}
\newcommand{\maxEpochs}{120}
\newcommand{\minEig}{0.068}
\newcommand{\mpEdge}{2.19}
\newcommand{\nBoot}{1000}
\newcommand{\nCapSettings}{9}
\newcommand{\nOpt}{26{,}569}

\newcommand{\pFrobHARtwo}{0.016}
\newcommand{\pFrobLSTMtwo}{\ensuremath{<}0.001}
\newcommand{\pHAR}{0.151}
\newcommand{\pHolmShHAR}{0.060}
\newcommand{\pHolmVolEVT}{0.021}
\newcommand{\pHolmVolEW}{0.010}
\newcommand{\pHolmVolSample}{0.105}
\newcommand{\pLSTM}{0.528}

\newcommand{\pabNoHawkes}{1.00}
\newcommand{\pabNoRMT}{1.00}
\newcommand{\pabNoTsallis}{1.00}
\newcommand{\parLSTM}{6512}
\newcommand{\parPINC}{1742}
\newcommand{\parPlain}{1712}
\newcommand{\patience}{15}
\newcommand{\pyVer}{3.11.14}

\newcommand{\qMax}{1.441}
\newcommand{\qMed}{1.330}
\newcommand{\qMin}{1.166}
\newcommand{\qlAlphaA}{0.3242}
\newcommand{\qlAlphaB}{0.3339}
\newcommand{\qlAlphaC}{0.3667}
\newcommand{\qlEVT}{0.3954}
\newcommand{\qlHAR}{0.3239}
\newcommand{\qlLSTM}{0.3186}
\newcommand{\qlLW}{0.4143}

\newcommand{\qlPINC}{0.3187}
\newcommand{\qlPlain}{0.3187}

\newcommand{\qlRMT}{0.3814}
\newcommand{\qlSample}{0.4924}

\newcommand{\qlsdPINC}{0.0008}
\newcommand{\ramGB}{36}
\newcommand{\rankPINCfive}{2}
\newcommand{\rankPINCthirty}{3}

\newcommand{\rmtPerm}{0.05}
\newcommand{\rmtShare}{0.387}
\newcommand{\rmtSigMax}{2}
\newcommand{\rmtSigMean}{1.10}
\newcommand{\seqLen}{10}

\newcommand{\shEVT}{1.57}
\newcommand{\shEVTzero}{1.65}

\newcommand{\shEWBHtwenty}{1.54}

\newcommand{\shHARtwenty}{0.20}

\newcommand{\shPINC}{1.19}
\newcommand{\shPINChi}{3.30}
\newcommand{\shPINClo}{\ensuremath{-}0.67}
\newcommand{\shPINCtwenty}{0.74}
\newcommand{\shPINCzero}{1.64}

\newcommand{\testEnd}{2015-01-06}
\newcommand{\testStart}{2014-02-06}
\newcommand{\testWin}{21}

\newcommand{\toEW}{0.006}

\newcommand{\toHAR}{0.218}

\newcommand{\toPINC}{0.152}
\newcommand{\toRMT}{0.035}
\newcommand{\torchVer}{2.13.0}
\newcommand{\trainDays}{189}

\newcommand{\turnLim}{0.6}
\newcommand{\valAlphaTen}{0.368}
\newcommand{\valAlphaZero}{0.330}
\newcommand{\valDays}{42}
\newcommand{\violEVaRhistOne}{9}
\newcommand{\violFivePINC}{14}
\newcommand{\violOnePINC}{5}

\newcommand{\volFull}{8.55}
\newcommand{\volHAR}{8.50}

\newcommand{\volNoRMT}{8.71}
\newcommand{\volPINC}{8.55}

\newcommand{\warmup}{63}

%% file: tab_data.tex
\begin{tabular}{@{}p{0.30\columnwidth}p{0.66\columnwidth}@{}}
\toprule
Item & Value \\
\midrule
Universe & \NStocks\ stocks  \\
Sample & 2012-12-05 to 2015-02-04; \NDays\ trading days, \NHalfDays\ half days \\
Bars & 1 min; 390 per full day (210 on half days); \NMinutes\ session minutes per stock \\
Gaps & strict intersection loses \PctLost\,\% of minutes; \PctImputed\,\% of ticker-minutes forward-filled ($\le 2$ min, within day) \\
Realised measures & 5-min sub-sampled RV and RC (five shifted grids) \\
Walk-forward & train \trainDays\ / validation \valDays\ / test \testWin\ days; \NFolds\ folds; \warmup-day warm-up \\
Out-of-sample & \NTestDays\ contiguous decision days, \testStart\ to \testEnd \\
Replication & \NSeeds\ seeds; \hpoTrials\ TPE trials per model and fold \\
\bottomrule
\end{tabular}

%% file: tab_forecast.tex
\begin{tabular}{@{}lrrrr@{}}
\toprule
Model & QLIKE$/N$ & s.d.\ seeds & Frobenius & Gauss.\ LL \\
\midrule
Sample & 0.4924 & -- & 10.81 & \ensuremath{-}40.09 \\
Ledoit--Wolf & 0.4143 & -- & 8.12 & \ensuremath{-}37.98 \\
RMT & 0.3814 & -- & 7.57 & \ensuremath{-}37.91 \\
EVT-GPD & 0.3954 & -- & 7.69 & \ensuremath{-}38.15 \\
HAR-RV & 0.3239 & -- & 6.30 & \ensuremath{-}36.53 \\
LSTM & \textbf{0.3186} & 0.0002 & 6.36 & \ensuremath{-}36.37 \\
PlainNN & 0.3187 & 0.0008 & 6.48 & \ensuremath{-}36.48 \\
PINC & 0.3187 & 0.0008 & 6.48 & \ensuremath{-}36.48 \\
\bottomrule
\end{tabular}

%% file: tab_dm.tex
\begin{tabular}{@{}lrrrr@{}}
\toprule
 & \multicolumn{2}{c}{QLIKE$/N$} & \multicolumn{2}{c}{Frobenius} \\
\cmidrule(lr){2-3}\cmidrule(lr){4-5}
PINC vs. & $\Delta$ & $p_{\rm Holm}$ & $\Delta$ & $p_{\rm Holm}$ \\
\midrule
Sample & \ensuremath{-}0.1737 & \ensuremath{<}0.001 & \ensuremath{-}4.337 & \ensuremath{<}0.001 \\
Ledoit--Wolf & \ensuremath{-}0.0955 & \ensuremath{<}0.001 & \ensuremath{-}1.643 & \ensuremath{<}0.001 \\
RMT & \ensuremath{-}0.0627 & \ensuremath{<}0.001 & \ensuremath{-}1.099 & \ensuremath{<}0.001 \\
EVT-GPD & \ensuremath{-}0.0766 & \ensuremath{<}0.001 & \ensuremath{-}1.211 & \ensuremath{<}0.001 \\
HAR-RV & \ensuremath{-}0.0052 & 0.151 & 0.174 & 1.000 \\
LSTM & 0.0001 & 0.528 & 0.119 & 1.000 \\
\bottomrule
\end{tabular}

%% file: tab_portfolio.tex
\begin{tabular}{@{}lrrrcrrr@{}}
\toprule
Strategy & Return (\%) & Vol.\ (\%) & Sharpe & 95\,\% CI & MaxDD (\%) & Turnover & $N_{\rm eff}$ \\
\midrule
EW-rebal & 15.4 & 10.08 & 1.53 & [\ensuremath{-}0.38, 3.67] & 6.70 & 0.006 & 29.0 \\
EW-BH & 15.6 & 10.10 & 1.54 & [\ensuremath{-}0.37, 3.68] & 6.73 & 0.000 & 28.8 \\
ERC-RMT & 14.6 & 9.68 & 1.50 & [\ensuremath{-}0.49, 3.69] & 6.41 & 0.011 & 27.5 \\
MV-Sample & 9.0 & 8.85 & 1.01 & [\ensuremath{-}0.93, 3.21] & 6.53 & 0.053 & 8.2 \\
MV-Ledoit--Wolf & 9.7 & 8.83 & 1.09 & [\ensuremath{-}0.87, 3.31] & 6.58 & 0.051 & 9.1 \\
MV-RMT & 10.0 & 8.69 & 1.15 & [\ensuremath{-}0.75, 3.38] & 6.55 & 0.035 & 7.3 \\
MV-EVT-GPD & 14.2 & 9.04 & \textbf{1.57} & [\ensuremath{-}0.29, 3.72] & \textbf{5.41} & 0.031 & 7.3 \\
MV-HAR-RV & 7.2 & \textbf{8.50} & 0.85 & [\ensuremath{-}0.95, 2.91] & 5.93 & 0.218 & 8.5 \\
MV-LSTM & 10.0 & 8.60 & 1.16 & [\ensuremath{-}0.66, 3.25] & 5.76 & 0.148 & 8.9 \\
MV-PlainNN & 10.2 & 8.55 & 1.19 & [\ensuremath{-}0.67, 3.30] & 5.61 & 0.152 & 8.8 \\
MV-PINC & 10.2 & 8.55 & 1.19 & [\ensuremath{-}0.67, 3.30] & 5.61 & 0.152 & 8.8 \\
EVaR-hist & 13.2 & 9.52 & 1.38 & [\ensuremath{-}0.46, 3.56] & 5.81 & 0.041 & 6.4 \\
\bottomrule
\end{tabular}

%% file: tabS_audit.tex
\begin{tabular}{@{}p{0.80\textwidth}c@{}}
\toprule
Requirement & Holds \\
\midrule
No look-ahead: realised date strictly after information date; target row t = day t+1 & yes \\
Train / validation / test strictly ordered; test days unique and contiguous & yes \\
Training independent of test data (poisoned-data leakage test, bit-identical) & yes \\
Features / estimators at t invariant to data after t (truncation tests) & yes \\
Scalers, HAR, Hawkes, EVT-threshold, $\alpha$, early stopping fitted without test data & yes \\
Hawkes = exact MLE on event times (numba / torch), recovered on simulated data & yes \\
HAR-RV = real daily/weekly/monthly OLS regression & yes \\
EVT = real GPD-POT MLE; fallbacks counted & yes \\
LSTM is a real nn.LSTM; PlainNN has equal capacity to PINC (+$\varphi$,$\theta$ only) & yes \\
Physics term is Fokker-Planck-INSPIRED (not a PDE residual); model named PINC & yes \\
$\hat\Sigma$ positive definite by construction; all forecasts finite & yes \\
Constraints (long-only, $\Sigma$w=1, cap, turnover) hold every day for every strategy & yes \\
Solver failures counted, never silent & yes \\
EVaR is an exponential-cone program, depends on $\alpha$, $\neq$ min-variance (controlled + real) & yes \\
Transaction cost kept separate from the turnover penalty (ridge) and the hard limit & yes \\
VaR/ES/EVaR forecasts are ex-ante (backcast / past covariance only); no full-sample thresholds & yes \\
MPS/CPU device chosen by measurement; MPS used only where faster & yes \\
torch.compile tested, enabled only if beneficial & yes \\
HPO budgets equal across LSTM and PlainNN; PINC reuses PlainNN hyper-parameters & yes \\
PINC differs from PlainNN in this run & no \\
RUN-LEVEL: full mode (not fast\_mode) & yes \\
RUN-LEVEL: $\geq$ 3 seeds for neural models & yes \\
RUN-LEVEL: all walk-forward folds used & yes \\
\bottomrule
\end{tabular}